# Physical properties of predicted MAX phase borides Hf$_2$AB (A = Pb, Bi): a DFT insight

M. S. Hossain, M. A. Ali*, M. M. Hossain, and M. M. Uddin

Department of Physics, Chittagong University of Engineering and Technology (CUET), Chattogram-4349, Bangladesh

**ABSTRACT**

We have used density functional theory to study the recently predicted MAX phase borides Hf$_2$AB (A = Pb, Bi) in where the mechanical, electronic, thermal, and optical properties have been investigated for the first time. A good agreement of the obtained lattice constants with the reported values confirmed the well accuracy of the present calculations. The stiffness constants ($C_{ij}$) attest the mechanical stability of all title compounds. The mechanical behaviors have been scrutinized discreetly by considering the bulk modulus, shear modulus, Young's modulus, as well as hardness parameters. The brittle nature of Hf$_2$AB (A = Pb, Bi) borides has also been confirmed. The electronic band structure and density of states (DOS) revealed the metallic behavior of the titled materials. The anisotropy in electrical conductivity has been disclosed by considering the energy dispersion along different directions. The variation of Vickers hardness is explained in terms of total DOS of Hf$_2$AB (A = Pb, Bi). The anisotropic nature of mechanical properties of the phases has also been studied. The technologically important parameters (Debye temperature, minimum thermal conductivity, and Grüneisen parameter) have also been used to evaluate the thermal behaviors of the titled materials. The possibility of Hf$_2$AB (A = Pb, Bi) for use as coating materials has been assessed by studying the reflectivity.

## 1. Introduction

The prospective demand for the chemical diversity of MAX phases is growing in time due to their increasing applications in many sectors. The list of applications is also being longer because of the beneficial properties of metals and ceramics [1–4]. A large number of articles have been published regularly by reporting either their new member or opening the new door for possible applications. The driving force that encouraged scientists is the bridging of their properties such as their conductivity (both electrical and thermal), machinability, and mechanical strength like metallic materials. On the other hand, their mechanical properties are good at high-temperature, their resistance to oxidation and corrosion is also high enough like ceramic materials [2,5–9]. The unique bridging of the aforementioned properties makes MAX phases one of the vital classes of materials and the research interest in these materials has been reflected from the exponential increase of the published articles [10]. To date, the number of known MAX phases crossed 150 mile-stone [11].

*Corresponding author: ashrafphy31@cuet.ac.bd

One of the limitations of MAX phases is that their diversity regarding X elements is confined within C and N for a long period as compounds [12–21] or solid solutions [22–31]. Thus, the extension of X elements beyond C/N can open the platform for a huge number of MAX phase members. The researchers have already found room for this extension by extending the X element to B (termed as MAX phase borides) owing to the bright prospect of B and its compounds [32]. It is now proven that the MAX phase borides are found as potential materials in many attractive fields [33–35]. Few numbers of MAX phase borides have also been successfully synthesized that make sure the possible extension of MAX phases in terms of borides. Rackl*et al.*[36,37] have synthesized the $M_2SB$ (Zr, Hf and Nb) MAX phase borides, they were further investigated by Ali *et al.*[38] using first principles calculations. Another important MAX phase borides (212 MAX phase: $Ti_2InB_2$) has been synthesized by Wang et al. [39]and they were further investigated by Ali *et al.*[40] and Wang *et al.*[41]. Moreover, a number of MAX phase borides have also been predicted by the researchers. Khazaei*et al.*[42] have predicted the MAX phase borides for the first time and reported the formation energies, electronic and mechanical properties of $M_2AlB$ (M = Sc, Ti, Cr, Zr, Nb, Mo, Hf, or Ta) MAX phase borides. Later on, the $Ti_2SiB$ [43], $M_2AB$ (M = Ti, Zr, Hf; A = Al, Ga, In) [44], $V_2AlB$ [10], and $M_2AlB$ (M = V, Nb, Ta) [45] have been predicted theoretically and investigated their physical properties as well.

The last inclusion in this MAX phase group (MAX phase borides) is being reported by Miao *et al.*[46]. Recently, Miao *et al*[46] have been predicted $Hf_2AB$ (A = Pb, Bi) borides by confirming their thermo-dynamical stability. They have also studied the electronic properties including band structure, the density of states and electron localization function. As we know that the MAX phase materials have been found themselves in many sectors for practical applications and this limited information is not enough to predict whether these borides have also the potential for applications in the fields where the MAX phase carbides or nitrides are already been used. To check this possibility, the physical properties of $Hf_2AB$ (A = Pb, Bi) borides should be compared with their corresponding carbides and/or nitrides. It is also an attractive topic that whether the borides MAX phase can be the replacement of corresponding carbides and/or nitrides.

Therefore, in our current project, we have investigated the mechanical, electronic, thermal and optical properties along with theoretical Vickers hardness and mechanical anisotropy of $Hf_2AB$ (A = Pb, Bi) borides for the first time. Moreover, the obtained physical properties of $Hf_2AB$ (A = Pb, Bi) are compared with those of other Hf-based borides $Hf_2AB$ (A = Al, Ga, In, S), where available.

## 2. Computational methodology

The physical properties of $Hf_2AB$ (A = Pb, Bi) borides are computed via plane-wave pseudopotential based density functional theory as implemented in the CAmbridge Serial Total Energy Package (CASTEP) code [47,48]. The calculations setting are given below: The exchange and correlations terms: generalized gradient approximation (GGA) of the Perdew–Burke–Ernzerhof (PBE) [49]; cutoff energy: 500 eV; k-point [50] mesh: 10 × 10 × 3; minimization technique: Broyden Fletcher Goldfarb Shanno (BFGS) [51]; electronic structure calculation: density mixing; energy convergence tolerance: $5 \times 10^{-6}$ eV/atom; maximum force tolerance: 0.01 eV/Å; maximum ionic displacement tolerance: $5 \times 10^{-4}$ Å; and maximum stress tolerance: 0.02 GPa. The pseudo-atomic calculations were performed for B - $2s^2 2p^1$, Pb- $5d^{10} 6s^2 6p^2$, Bi - $6s^2 6p^3$, and Hf - $5d^2 6s^2$ electronic orbitals.

## 3. Results and discussion

### 3.1. *Structural properties*

The characteristic unit cell for 211 MAX phases is shown in Fig. 1, crystallized with hexagonal structure [Space groupP63/mmc][1]. The $Hf_2AB$ (A = Pb, Bi) borides belong to the 211 sub-class of MAX phases, therefore, the $Hf_2AB$ (A = Pb, Bi) has the same unit cell characteristics except for the difference in the lattice constants and internal parameter. The atomic positions of Hf, Pb/Bi and B atoms in the unit cell are (1/3, 2/3, $z_M$), (1/3, 2/3, 3/4) and (0, 0, 0), respectively. The values of the internal parameter, $z_M$ are given in Table 1. The optimized lattice constants (*a*, *c*), internal parameter and *c/a* ratio of$Hf_2AB$ (A = Pb and Bi) are listed in Table 1 along with the reported results for comparison. As seen in Table 1 that the calculated values of *a* and *c* are very close to the reported values. A very low deviation from the available theoretical values certifies the accuracy of the present calculations. No experimental values are available for comparison since these phases are not synthesized yet.

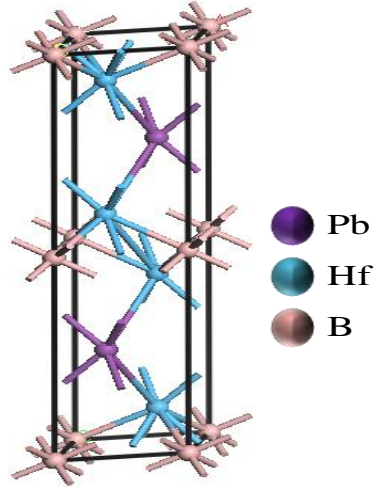

**Fig. 1 -** Crystal structure of the Hf$_2$PbB compound.

**Table 1** Calculated lattice parameters (*a* and *c*), internal parameter ($z_M$) and *c/a* ratio of Hf$_2$AB (A = Pb, Bi) MAX phases.

| Phase | *a* (Å) | % of deviation | *c* (Å) | % of deviation | $z_M$ | *c/a* | Reference |
|---|---|---|---|---|---|---|---|
| Hf$_2$PbB | 3.493 | 0.721 | 14.899 | 0.454 | 0.0840 | 4.267 | This work |
|  | 3.468 |  | 14.967 |  | 0.0826 | 4.316 | [46] |
| Hf$_2$BiB | 3.530 | 0.199 | 14.374 | 0.794 | 0.0867 | 4.07 | This work |
|  | 3.523 |  | 14.489 |  | 0.0826 | 4.113 | [46] |

## 3.2 Mechanical properties

To explore the mechanical behavior of Hf$_2$AB (A = Pb, Bi) MAX phases, we have calculated the mechanical properties characterizing parameters using the strain-stress method [41,52–54] as implement in the CASTEP code. Before starting with the mechanical properties, the mechanical stability of $Hf_2AB$ (A = Pb, Bi) should be checked by employing well known stability conditions for the hexagonal system based on the stiffness constants ($C_{ij}$). The obtained stiffness constants satisfies the following conditions [55,56]: $C_{11} > 0$, $C_{33} > 0$, $C_{44} > 0$, $C_{11}$-$C_{12} > 0$, $(C_{11} + C_{12})C_{33} - 2C_{13}^2 > 0$; hence the $Hf_2AB$ (A = Pb, Bi) MAX members are mechanically stable. The obtained stiffness constants are compared with those of other Hf-based [$Hf_2AB$ (A = Al, Ga, In, S)] borides as shown in Fig. 2 (a). It is seen that the stiffness constants of $Hf_2AB$ (A = Pb, Bi) are lower than those of $Hf_2AB$ (A = Al, S) borides but higher than those of $Hf_2AB$ (A = Ga, In) borides. The obtained values are also presented in Table 2 for an easy understanding of the exact values. It is well known that the $C_{11}$ and $C_{33}$ measure the stiffness of solids along *a* and *c*-axis, therefore, the differences in the values of $C_{11}$ and $C_{33}$ indicating the anisotropic nature of the bonding strength. It is also observed that for the $Hf_2PbB$, $C_{11} > C_{33}$, indicating stiffer nature along *a*-axis compared to *c*-axis like $Hf_2AB$ (A = Ga, In) borides whereas for the $Hf_2BiB$, $C_{11} < C_{33}$, indicating stiffer nature along the *c*-axis compared to *a*-axis like $Hf_2AB$ (A = Al, S) borides [38,44]. Moreover, these stiffness constants are further used to calculate the polycrystalline elastic moduli using well-known formalisms. For example, the bulk modulus (*B*) and shear modulus (*G*) are calculated using Hill's approximation [57,58] that usually expressed the average values of the upper limit (Voigt [59]) and lower limit (Reuss [60]) of *B*: [$B = (B_V + B_R)/2$] and *G*: [$G = (G_V + G_R)/2$]. The $B_V$, $B_R$, $G_V$ and $G_R$ are calculated from $C_{ij}$ using the following equations: $B_V = [2(C_{11} + C_{12}) + C_{33} + 4C_{13}]/9$ ; $B_R = C^2/M$; $C^2 = C_{11} + C_{12})C_{33} - 2C_{13}^2$; $M = C_{11} + C_{12} + 2C_{33} - 4C_{13}$; $G_V = [M + 12C_{44} + 12C_{66}]/30$ and $G_R = \left(\frac{5}{2}\right)[C^2 C_{44} C_{66}]/[3B_V C_{44} C_{66} + C^2(C_{44} + C_{66})]$; $C_{66} = (C_{11} - C_{12})/2$. The Young's modulus and Poisson's ratio are also calculated from the *B* and *G* using the relations: $Y = 9BG/(3B + G)$ and $\upsilon = (3B - Y)/(6B)$ [61,62]. The elastic moduli (*B, G, Y*) of $Hf_2AB$ (A = Pb, Bi) are higher than those of $Hf_2AB$ (A = Ga, In) but lower than those of $Hf_2AB$ (A = Al, S); are not directly related to the hardness of solids but these moduli are higher for harder solids. Therefore, the hardness of $Hf_2AB$ (A = Pb, Bi) is expected to be lower than that of $Hf_2AB$ (A = Al, S) but higher than that of $Hf_2AB$ (A = Ga, In). One of the ways of comparing hardness is the calculations of hardness parameters $H_{macro}$ and $H_{micro}$ for compounds having a similar structure [63].

**Table 2** The elastic constants, $C_{ij}$ (GPa), bulk modulus, $B$ (GPa), shear modulus, $G$ (GPa), Young's modulus, $Y$ (GPa), micro hardness, $H_{micro}$ (GPa), macro hardness, $H_{macro}$ (GPa), Pugh ratio, $G/B$, and Poisson ratio, $v$, of $Hf_2AB$ (A = Pb, Bi), together with those of $Hf_2AB$ (A = Al, Ga, In, S) for comparison.

| Phase | $C_{11}$ | $C_{12}$ | $C_{13}$ | $C_{33}$ | $C_{44}$ | $B$ | $G$ | $Y$ | $H_{macro}$ | $H_{micro}$ | $G/B$ | $v$ | Reference |
|---|---|---|---|---|---|---|---|---|---|---|---|---|---|
| $Hf_2PbB$ | 225 | 58 | 51 | 206 | 75 | 108 | 76 | 192 | 15.01 | 15.63 | 0.73 | 0.21 | This work |
| $Hf_2BiB$ | 206 | 66 | 70 | 212 | 82 | 115 | 72 | 184 | 11.94 | 13.20 | 0.67 | 0.23 | This work |
| $Hf_2AlB$ | 232 | 72 | 81 | 267 | 109 | 133 | 92 | 223 | 15.31 | 17.14 | 0.69 | 0.22 | [44] |
| $Hf_2GaB$ | 213 | 77 | 63 | 176 | 66 | 111 | 67 | 167 | 9.96 | 11.20 | 0.60 | 0.25 | [44] |
| $Hf_2InB$ | 210 | 69 | 50 | 197 | 55 | 106 | 65 | 163 | 9.97 | 11.11 | 0.61 | 0.25 | [44] |
| $Hf_2SB$ | 285 | 81 | 92 | 298 | 130 | 156 | 112 | 271 | 18.45 | 21.63 | 0.72 | 0.21 | [38] |

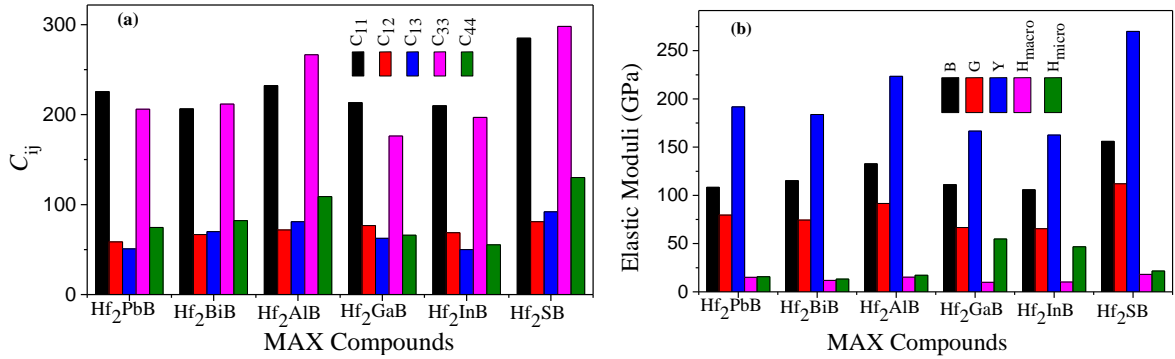

**Fig. 2** Comparison of (a) $C_{ij}$ and (b) $B$, $G$, $Y$, $H_{macro}$ and $H_{micro}$ of $Hf_2AB$ (A = Pb, Bi) with those of $Hf_2AB$ (A = Al, Ga, In, S).

To do this, we have used the elastic moduli further to calculate the hardness parameters using the equations: $H_{macro} = 2[(\frac{G}{B})^2 G]^{0.585} - 3$ [64,65] and $H_{micro} = \frac{(1-2v)Y}{6(1+v)}$ [66]. These two equations expressed the role of elastic moduli to the hardness of solids; $G$ is more related to the $H_{macro}$ compared to $B$ while $Y$ is directly related to the $H_{micro}$. Fig. 2 (b) shows a comparison among the elastic moduli and hardness parameters of $Hf_2AB$ (A = Pb, Bi) together with those of $Hf_2AB$ (A = Al, Ga, In, S) MAX phase borides. Like $C_{ij}$, the elastic moduli and hardness parameters of $Hf_2AB$ (A = Pb, Bi) are lower than those of $Hf_2AB$ (A = Al, S) borides but higher than those of $Hf_2AB$ (A = Ga, In) borides. The values are also presented in Table 2 for an easy understanding of the exact values. Though the hardness parameters are larger for the $Hf_2PbB$ compared to the $Hf_2BiB$ but the $C_{44}$ of $Hf_2BiB$ (82 GPa) is slightly larger than that of $Hf_2PbB$ (75 GPa). Among the mechanical properties characterizing parameters, $C_{44}$ is the best hardness predictor as suggested by Jhi et al.[67] based on the results for binary carbides and nitrides. Therefore, the hardness should be higher for the $Hf_2BiB$ compared to the $Hf_2PbB$. Actually, these hardness parameters do not predict exact hardness but efficient to predict comparative hardness based on the values of elastic moduli which are usually higher for harder solids. In order to predict the hardness more exactly, we have calculated the Vickers hardness ($H_v$) using the formula proposed by Gou et al.[68] for partial metallic solids. The detail of the formalism can be found elsewhere [69–72]. The important feature of this method is that it is the geometrical averages of all bonds present in the solids. The calculated Vickers hardness of $Hf_2AB$ (A = Pb, Bi) is presented in Table 3. As seen in Table 3, the $H_v$ of the $Hf_2BiB$ (2.85 GPa) is slightly higher than that of $Hf_2PbB$ (2.56 GPa), in good agreement with the Jhi et al.[67] statement.

**Table 3** Calculated Mulliken bond number $n^\mu$, bond length $d^\mu$, bond overlap population $P^\mu$, metallic population $P^{\mu'}$, bond volume $v_b^\mu$, bond hardness $H_v^\mu$ of $\mu$-type bond and Vickers hardness $H_v$ of $Hf_2AB$ (A = Pb, Bi).

| Compounds | Bond | $n^\mu$ | $d^\mu$ (Å) | $P^\mu$ | $P^{\mu'}$ | $v_b^\mu$ (Å$^3$) | $H_v^\mu$ (GPa) | $H_v$ (GPa) |
|---|---|---|---|---|---|---|---|---|
| $Hf_2PbB$ | Hf-B | 4 | 2.37382 | 1.68 | 0.1028 | 39.367 | 2.56 | 2.56 |
| $Hf_2BiB$ | Hf-B | 4 | 2.38886 | 1.73 | 0.0175 | 38.781 | 2.85 | 2.85 |

The difference in the values of Vickers hardness, in other words the bonding strength can be explain in terms of density of states (DOS) that will be presented in the 3.3 section.

*The brittleness of $Hf_2AB$ (A = Pb, Bi)*

The most interesting characteristic of MAX phases is the extraordinary combinations of properties of metals and ceramics [3]. Like metals, they are machinable as discussed in the previous section; they are also brittle like ceramic materials. The brittleness of $Hf_2AB$ (A = Pb, Bi) is evaluated by the Pugh ratio (G/B) and Poisson's ratio ($\upsilon$). The calculated values of G/B and $\upsilon$ are listed in Table 2. A critical value of 0.571 for G/B [73] is considered as a borderline for ductile to brittle transition whereas the critical value of $\upsilon$ is 0.26 [74] that fixed the brittle-ductile borderline. As seen in Table 2 that the G/B > 0.572 and $\upsilon$ < 0.26, thus, the $Hf_2AB$ (A = Pb, Bi) are fall into the brittle class of materials.

## 3.3 Electronic Properties

The detail of the electronic nature of $Hf_2AB$ (A = Pb, Bi) can be explained by exploring the electronic band structure and density of states (DOS) (total and partial). The electronic band structures of $Hf_2AB$ (A = Pb, Bi) is shown in Fig. 3 (a and b). The Fermi level is set at zero of the energy scale. As seen from the figures that no band gap is appeared owing to the overlapping of the valence band and conduction band. Consequently, metallic nature is exhibited by the MAX phase borides [$Hf_2AB$ (A = Pb, Bi)]. Moreover, the anisotropic nature in electronic conductivity is also exhibited by the band structure of $Hf_2AB$ (A = Pb, Bi) borides. The paths Γ-A, H-K and M-L exhibit the energy dispersion in the *c*-direction whereas the paths A-H, K-Γ, Γ-M and L-H exhibit energy dispersion in the basal planes. As observed from Fig 3, the degree of energy dispersion is lesser along the *c*-direction compared to the basal plane [75] owing to the higher effective mass tensor for conduction along *c*-direction than that of the basal plane (*ab*-plane) [76].

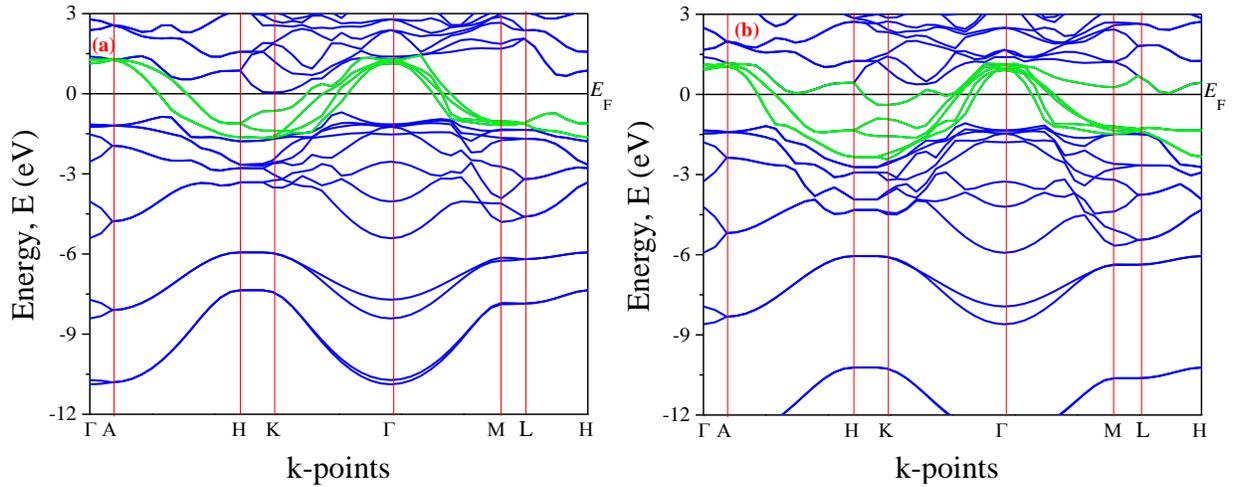

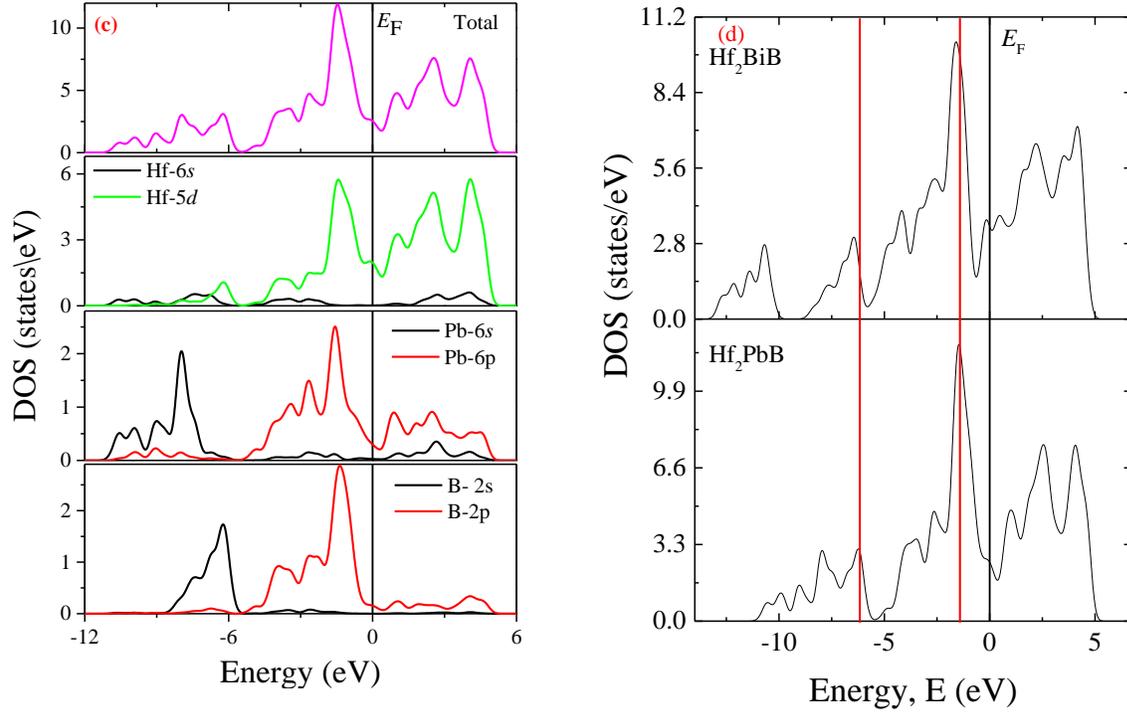

**Fig. 3** Electronic band structure (a) Hf$_2$PbB and (b) Hf$_2$BiB; (c) total and partial DOS of Hf$_2$PbB; (d) total DOS of Hf$_2$AB (A = Pb and Bi) MAX phases.

To evaluate the contribution from the different electronic states to the electronic conductivity we have also calculated the total and partial DOS of Hf$_2$AB (A = Pb, Bi) borides. Owing to the similar nature except for the differences in the position of the peaks [can be understood from the position of the curves in band structures shown in Fig. 3 (a and b)] only the total and partial DOS of Hf$_2$PbB is presented in Fig. 3 (c). As seen in the figure that the prime contribution comes from the Hf-5$d$ states to the DOS at the Fermi level while Pb-6$p$ and B-2$p$ also contribute but lesser in magnitude compared to the Hf-5$d$ orbital. The partial DOS is also an effective tool to disclose the hybridization among the different states. The valence band can be divided into energy ranges: - 12.0 eV to -6.0 eV and -6.0 eV to - 0.0 eV. The Pb-6$s$ and B-2$s$ hybridized strongly that lead the peaks in the low energy region of the total DOS. The energy region of 0 eV to -6 eV is dominated by Hf-5$d$ which is also hybridized with the Pb-6$p$ and B-2$p$.

The difference in the Vickers hardness can be explained with the help of DOS. For this reason, we have also calculated the DOS of Hf$_2$AB (A = Pb, Bi) as shown in Fig. 3 (d) in which a black solid line refers to the Fermi level and two solid red lines are used to demonstrate the shifting of peaks in the DOS. As observed in Fig. 3 (d), the positions of the peaks are slightly shifted to the left (low energy side) for the Hf$_2$BiB phase. The peaks in the DOS are the consequence of the hybridization between different orbitals and the peak's position defines the energy of the hybridized states. The energy measures the strength of the hybridization, consequently, the bonding strength between unlike atoms. In general, the lower position of the peaks, the higher strength of the bonds [76]. Therefore, stronger bonding between unlike atoms is expected for the Hf$_2$BiB compared to the Hf$_2$PbB as consistent from the values of $C_{44}$ (Table 2) and Vickers hardness (Table 3).

## 3.3 The elastic anisotropy

In Table 2, it is observed that $C_{11} > C_{33}$ for the Hf$_2$PbB and $C_{11} < C_{33}$ for the Hf$_2$BiB. These unequal values confirm the anisotropic nature of the mechanical properties because other parameters such as different polycrystalline elastic moduli are calculated from the stiffness constants $C_{ij}$. This aspect encouraged us to investigate mechanical anisotropy because some physical processes are greatly influenced by the anisotropic nature of elastic moduli [77]. Moreover, the study of anisotropy provides the information required to improve the stability of solids for many applications [78]. Thus, the anisotropic nature of Hf$_2$AB (A = Pb, Bi) is investigated considering different formalisms. At first, we investigate the different anisotropic factors directly related to the stiffness constants. For example, the shear anisotropic factors for the {100}, {010} and {001} planes is computed using the equations: $A_1 = \frac{1/6(C_{11}+C_{12}+2C_{33}-4C_{13})}{C_{44}}, A_2 = \frac{2C_{44}}{C_{11}-C_{12}}, A_3 = A_1 \cdot A_2 = \frac{1/3(C_{11}+C_{12}+2C_{33}-4C_{13})}{C_{11}-C_{12}}$ [79], respectively and presented in Table 4. As observed in Table 4, the values of $A_i$'s are not equal to 1 (one). $A_i = 1$, implies the isotropic nature, thus, the compounds in interest are anisotropic ($A_i \neq 1$). The unequal values of bulk modulus along $a$ and $c$-axis are estimated by the following equations [80]: $B_a = a\frac{dP}{da} = \frac{\Lambda}{2+\alpha}, B_c = c\frac{dP}{dc} = \frac{B_a}{\alpha}$, where $\Lambda = 2(C_{11} + C_{12}) + 4C_{13}\alpha + C_{33}\alpha^2$ and $\alpha = \frac{(C_{11}+C_{12})-2C_{13}}{C_{33}+C_{13}}$. The values of $B_a$ and $B_c$ are not equal, which implies the anisotropy of Hf$_2$AB (A = Pb, Bi). The linear compressibility ($k$) are different along the $a$ and $c$-axis. Their ratio is estimated by the following relation[81]: $\frac{k_c}{k_a} = C_{11} + C_{12} - 2C_{13}/(C_{33} - C_{13})$. A value of $k_c/k_a=$

1, implies the isotropic nature of solids, thus, $k_c/k_a \neq 1$(Table 4) for compounds in interests indicating the anisotropy of Hf$_2$AB (A = Pb, Bi).

**Table 4** Anisotropic factors, $A_1$, $A_2$, $A_3$, $k_c/k_a$, $B_a$, $B_c$, percentage anisotropy factors $A_G$ and $A_B$, and universal anisotropic index $A^U$ of Hf$_2$AB (A = Pb, Bi), together with those of the Hf$_2$AB (A = Al, Ga, In and S) for comparison.

| Phase | $A_1$ | $A_2$ | $A_3$ | $B_a$ | $B_c$ | $k_c/k_a$ | $A_B$ | $A_G$ | $A^U$ | References |
|---|---|---|---|---|---|---|---|---|---|---|
| Hf$_2$PbB | 1.10 | 0.89 | 1.01 | 301 | 424 | 1.172 | 0.156 | 0.163 | 0.019 | This work |
| Hf$_2$BiB | 0.84 | 1.17 | 0.99 | 293 | 620 | 0.94 | 0.017 | 0.322 | 0.033 | This work |
| Hf$_2$AlB | 0.79 | 1.36 | 1.07 | 366 | 479 | 0.76 | 0.021 | 0.017 | 0.480 | [44] |
| Hf$_2$GaB | 0.98 | 0.97 | 0.96 | 381 | 263 | 1.45 | 0.023 | 0.018 | 0.572 | [44] |
| Hf$_2$InB | 1.43 | 0.78 | 1.12 | 340 | 279 | 1.22 | 0.024 | 0.023 | 0.711 | [44] |
| Hf$_2$SB | 0.76 | 1.27 | 0.97 | 447 | 506 | 0.88 | 0.067 | 0.788 | 0.080 | [38] |

Anisotropy is attributed due to the differences in the values of bulk modulus and shear modulus obtained using Voigt and Reuss models. This anisotropy is calculated using the formulae: $A_B = \frac{B_V - B_R}{B_V + B_R} \times 100\%$ and $A_G = \frac{G_V - G_R}{G_V + G_R} \times 100\%$ [82]. The universal anisotropic index $A^U$ is computed using the equation based on the obtained values of $B$ and $G$ using Voigt and Reuss models [83]: $A^U = 5\frac{G_V}{G_R} + \frac{B_V}{B_R} - 6 \geq 0$. The zero values of $A_B$, $A_G$ and $A^U$ define the solid as isotropic otherwise anisotropic materials. Thus, the non-zero values of the mentioned parameters (Table 4) imply that the compounds under consideration are anisotropic.

In addition, we have also investigated the anisotropy in elastic moduli of Hf$_2$AB (A = Pb, Bi) such as Young's modulus ($Y$), linear compressibility ($K=1/B$), shear modulus ($G$) and Poisson's ratio ($v$) by presenting their values in 2D and 3D. The 2D and 3D plots are presented in Fig. 4(a-d) for the Hf$_2$PbB and Fig. 5 (a-d) for the Hf$_2$BiB phase that have been created using the ELATE code [84,85] based on their stiffness matrix.

(a) Young's modulus (*Y*),

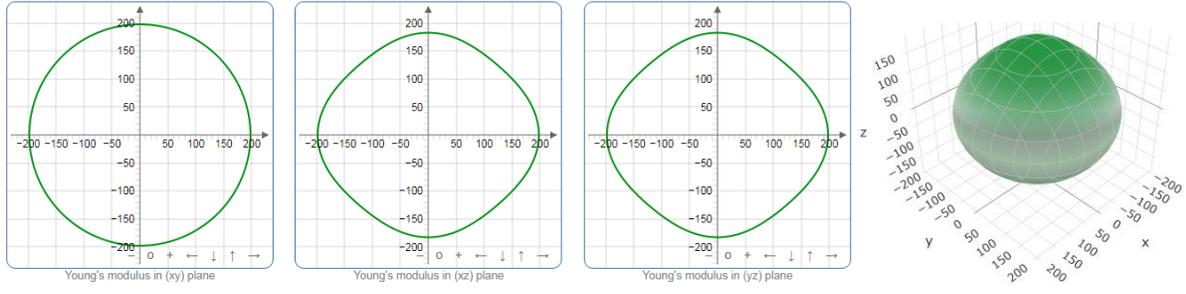

(b) Linear compressibility (*K*=1/*B*)

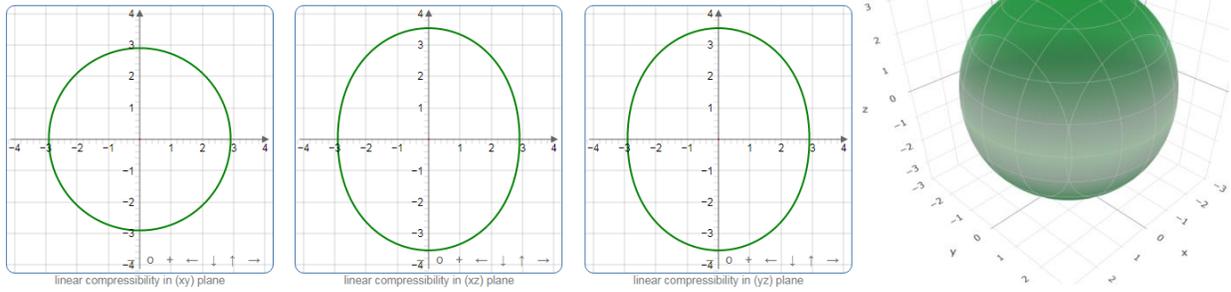

(c) Shear modulus (*G*)

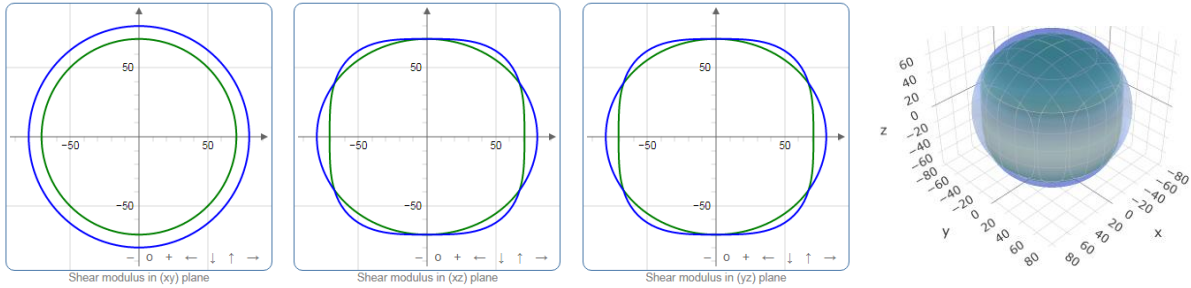

(d) Poisson's ratio (*υ*)

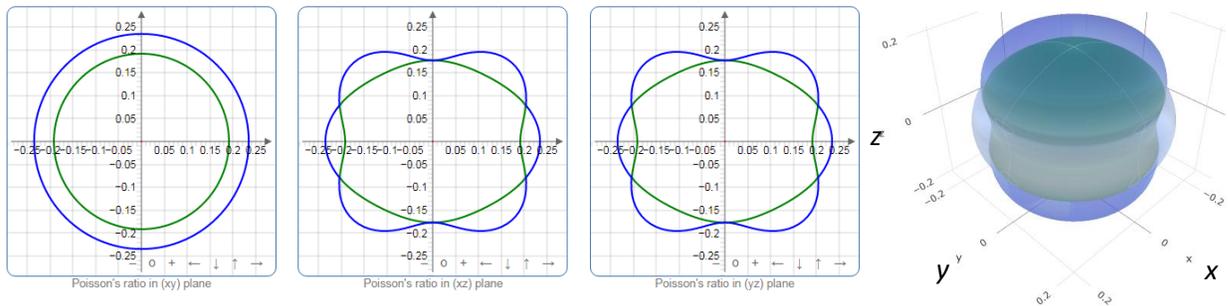

**Fig. 4:** The 2D and 3D plots of (a) *Y*, (b) *K*, (c) *G* and (d) *υ* of the Hf$_2$PbB compound.

(a) Young's modulus (Y),

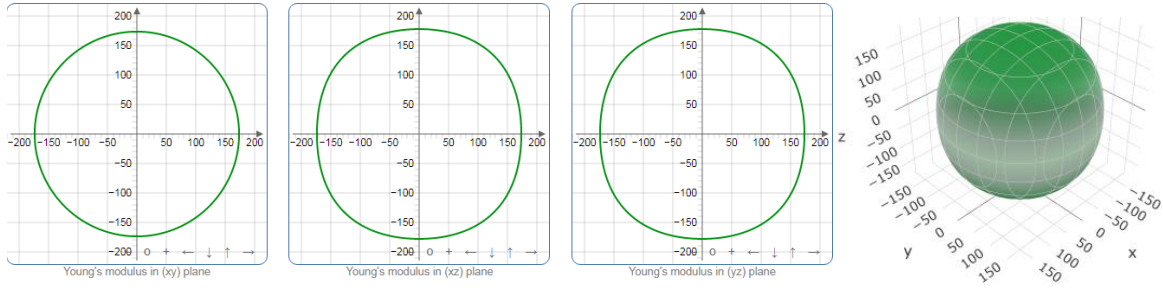

(b) Linear compressibility (K=1/B)

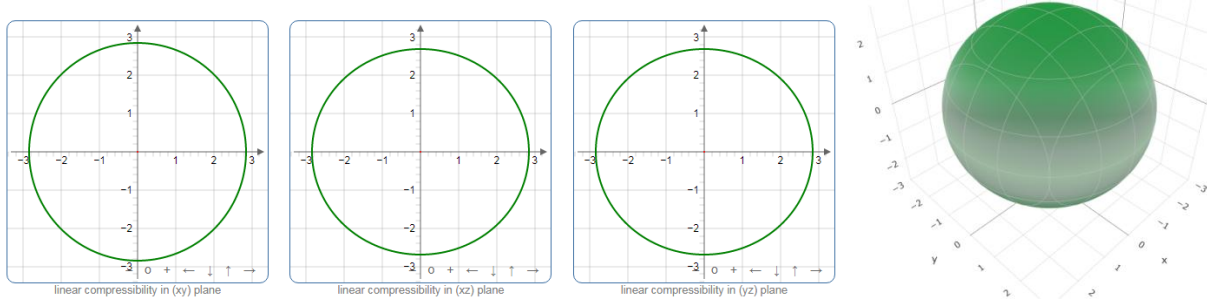

(c) Shear modulus (G)

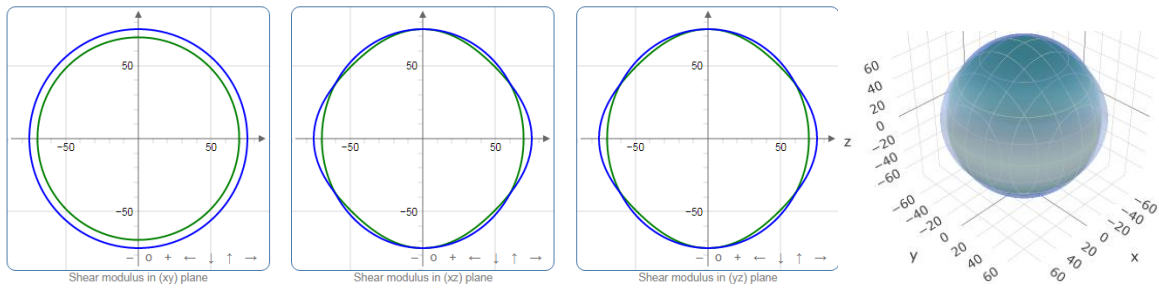

(d) Poisson's ratio (v)

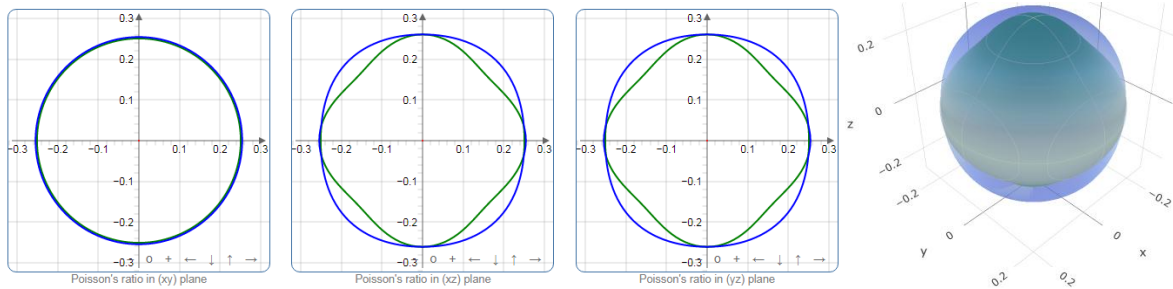

**Fig. 5:** The 2D and 3D plots of (a) $Y$, (b) $K$, (c) $G$ and (d) $v$ of the Hf$_2$BiB compound.

A perfect circle/sphere represents an isotropic solids and vice versa. The anisotropy level can be understood from the deviation of circular (for 2D) and spherical (for 3D) presentation of the elastic moduli. For example, comparatively a greater extent of deviation from circle or sphere is observed for the Hf$_2$PbB compared to Hf$_2$BiB, suggesting a more anisotropic nature as seen from values of $A_Y$, $A_K$, $A_G$, $A_\upsilon$ in Table 5. 2D plots of elastic moduli help to understand the anisotropic nature exhibited by 3D plots. For example, it is seen in 2D plots of Y that the Hf$_2$PbB phase has a maximum value of Y at the horizontal axis in both $xz$ and $yz$ planes which is to be isotropic in $xy$ plane while the Hf$_2$BiB phase has the maximum value of Y at an angle of 45° in between the axes. The compressibility is maximum along the vertical axis of both $xz$ and $yz$ planes for the Hf$_2$PbB while it is maximum along the horizontal axis in both $xz$ and $yz$ planes for the Hf$_2$BiB. The shear modulus is minimum at both axes in both $xz$ and $yz$ planes for the Hf$_2$PbB while it is maximum at the vertical axes in both $xz$ and $yz$ planes for the Hf$_2$BiB. The Poisson's ratio also exhibits different anisotropy for both compounds. It is minimum at both axes in both $xz$ and $yz$ planes for the Hf$_2$PbB phase but maximum at both axes in both $xz$ and $yz$ planes for the Hf$_2$BiB phase. A common thing is present for the compounds; all the elastic moduli are to be isotropic in $xy$ plane.

**Table 5:** The minimum and the maximum values of the Young's modulus, compressibility, shear modulus, and Poisson's ratio of Hf$_2$AB (A = Pb, Bi).

| Phase | $Y_{min.}$ (GPa) | $Y_{max.}$ (GPa) | $A_Y$ | $K$ $(TPa^{-1})$ | $K$ $(TPa^{-1})$ | $A_K$ | $G_{min.}$ (GPa) | $G_{max.}$ (GPa) | $A_G$ | $\upsilon_{min.}$ | $\upsilon_{max.}$ | $A_\upsilon$ |
|---|---|---|---|---|---|---|---|---|---|---|---|---|
| Hf$_2$PbB | 182.95 | 203.05 | 1.11 | 2.909 | 3.419 | 1.17 | 74.66 | 83.49 | 1.12 | 0.179 | 0.233 | 1.30 |
| Hf$_2$BiB | 173.01 | 193.07 | 1.12 | 2.774 | 2.952 | 1.06 | 69.58 | 82.19 | 1.18 | 0.174 | 0.273 | 1.57 |

*3.3 Thermal properties*

MAX phases have good mechanical properties at high temperature that makes them potentials candidates for high-temperature technology. Thus, the dissemination of knowledge regarding the high-temperature application is of scientific interest. This can be done by evaluating thermal properties characterizing parameters such as Debye temperature ($\Theta_D$), minimum thermal conductivity ($K_{min}$) and Grüneisen parameter ($\gamma$) and so on. In this study, we have calculated $\Theta_D$, $K_{min}$ and $\gamma$ of Hf$_2$AB (A = Pb, Bi) and tried to correlate with the mechanical properties for the same. The thermal properties characterizing parameters of Hf$_2$AB (A = Al, Ga, In) are also calculated using published data. The Debye temperature ($\Theta_D$) is considered as a fundamental thermal parameter that can link the mechanical properties of solids with other important physical processes [70,76,86]. Anderson's method [87] is widely used [63,69,70,72,88] and very simple to calculate the $\Theta_D$ by calculating the sound velocities passing through the solids. The equation used in this calculation is the following: $\Theta_D = h/k_B \left[ (3n/4\pi) N_A \rho/M \right]^{1/3} v_m$, where, $M$ be the molar mass, $n$ be the number of atoms in the molecules, $\rho$ be the mass density, and $h$, $k_B$, and $N_A$ be the Plank's constant, Boltzmann constant and Avogadro's number, respectively. The $v_m$ is the mean sound velocity that is obtained by the formula: $v_m = \left[ 1/3 \left( 1/v_l^3 + 2/v_t^3 \right) \right]^{-1/3}$ where, $v_l$ and $v_t$ are the velocity of sound for longitudinal and transverse mode, respectively. The $v_l$ and $v_t$ are estimated using the following equations: $v_l = \left[ (3B + 4G)/3\rho \right]^{1/2}$ and $v_t = \left[ G/\rho \right]^{1/2}$. The obtained values of density, longitudinal, transverse and average sound velocities ($v_l$, $v_t$, and $v_m$, respectively) and the $\Theta_D$ of $Hf_2AB$ (A = Pb, Bi) are presented in Table 6. The $\Theta_D$ of $Hf_2AB$ (A = Pb, Bi) is close to each other like their mechanical properties characterizing parameters. Though the Vickers hardness of the $Hf_2BiB$ is lightly higher than that of $Hf_2PbB$, thus, higher $\Theta_D$ for the $Hf_2BiB$ compound is expected. But, the $\Theta_D$ for the $Hf_2BiB$ is slightly smaller than that of the $Hf_2PbB$. The reason might be that the atomic mass of Bi is higher than that of Pb that leads to a higher density of the $Hf_2BiB$, results a lower $\Theta_D$ value for the $Hf_2BiB$ compared to $Hf_2PbB$ [72]. The inverse role of density can be seen in the formulae used to calculate the $\Theta_D$. The values of $v_l$, $v_t$, and $v_m$ are also in agreement with our explanation. The $\Theta_D$ values for other Hf-based 211 borides presented here are in well consistent with our assumption.

Minimum thermal conductivity ($K_{min}$) becomes one of the fundamental parameters for materials which are used at high-temperature application, thus, the study of $K_{min}$ of MAX phase materials is of scientific importance because of their usefulness in the high-temperature technology. The thermal conductivity of materials (e.g., ceramics) downs to a constant value ($K_{min}$) at high temperatures. The $K_{min}$ of $Hf_2AB$ (A = Pb, Bi) has been estimated using the equation [89]: $K_{min} = k_B v_m \left( \frac{M}{n\rho N_A} \right)^{-2/3}$, where $k_B$, $v_m$, $N_A$ and $\rho$ are Boltzmann constant, average phonon velocity, Avogadro's number and density of crystal, respectively and presented in Table 6. The reported values of $K_{min}$ for the compounds are also given in Table 6 that are comparable with that of other MAX phases [30,90]. The values are close to each other for the compound, revealing the usefulness of $Hf_2AB$ (A = Pb, Bi) as an alternative of $Hf_2AB$ (A = Al, Ga, In and S) compounds for desired applications.

**Table 6** Calculated density ($\rho$), longitudinal, transverse and average sound velocities ($v_l$, $v_t$, and $v_m$, respectively), Debye temperature ($\Theta_D$), minimum thermal conductivity ($K_{min}$) and Grüneisen parameter ($\gamma$) of Hf$_2$AB (A = Pb, Bi) borides, together with those of the Hf$_2$AB (A = Al, Ga, In and S) for comparison.

| Phase | $\rho$ (g/cm$^3$) | $v_l$ (m/s) | $v_t$ (m/s) | $v_m$ (m/s) | $\Theta_D$ (K) | $K_{min}$ (W/mK) | $\gamma$ | Reference |
|---|---|---|---|---|---|---|---|---|
| Hf$_2$PbB | 12.12 | 4145 | 2504 | 2768 | 305 | 0.796 | 1.19 | This work |
| Hf$_2$BiB | 12.33 | 4175 | 2416 | 2681 | 297 | 0.778 | 1.33 | This work |
| Hf$_2$AlB | 9.10* | 5301* | 3180* | 3518* | 399* | 0.866# | 1.49* | This work |
| Hf$_2$GaB | 10.26* | 4417* | 2557* | 2836* | 324* | 0.713# | 1.36* | This work |
| Hf$_2$InB | 10.43* | 4297* | 2496* | 2769* | 308* | 0.661# | 1.50* | This work |
| Hf$_2$SB | 10.19 | 5465 | 3309 | 3657 | 430 | 0.792 | 1.50 | [38] |

*Calculated using published data, #Ref-[44]

The Grüneisen parameter ($\gamma$) is an important thermal parameter from which the anharmonic effects owing to the lattice dynamics can be understood. We have calculated the $\gamma$ for Hf$_2$AB (A = Pb, Bi) compounds using the following relationship between $\gamma$ and $\upsilon$[91]: $\gamma = \frac{3}{2}\frac{(1+\nu)}{(2-3\nu)}$. The $\gamma$ for Hf$_2$AB (A = Al, Ga and In) have also been calculated using the published data for comparison. As observed in Table 6, the computed values of $\gamma$ are found to be in the range of .85 to 3.53 as expected for the polycrystalline materials having $\upsilon$ in the range of 0.05–0.46 [92]. Besides, quite low values of $\gamma$ certify the low anharmonic effects within the compounds in interest. Moreover, the anharmonic effect is lesser in Hf$_2$AB (A = Pb, Bi) compounds compared to Hf$_2$AB (A = Al, Ga, In and S) borides.

Although, the thermal expansion coefficient is still unknown but due to the low $K_{min}$ and considerably high melting temperature of Hf$_2$AB (A = Pb, Bi), it is expected that they could be potential candidates for thermal barrier coating (TBC) materials in high temperature technology.

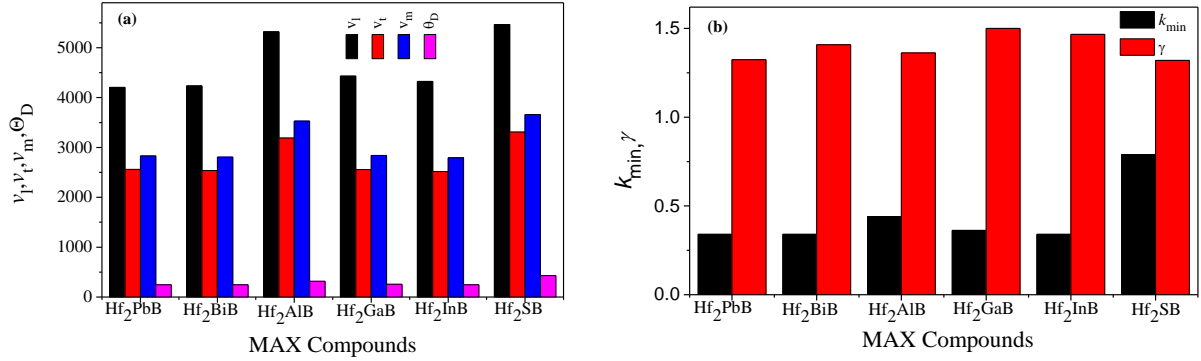

**Fig. 6** Comparison of (a) sound velocities and Debye temperature and (b) minimum thermal conductivity and Grüneisen parameter of $Hf_2AB$ (A = Pb, Bi) with those of $Hf_2AB$ (A = Al, Ga, In, S).

### 3.6 *Assessment of suitability as coating materials*

One of the potential applications of the MAX phase materials is the use as outside layer materials to protect the inside from solar radiation. This sub-section is dedicated to evaluate the interaction between the electromagnetic (EM) radiation and studied materials. The interactions of EM ray and materials exhibit two physical phenomena: energy transformation (absorption, reflectivity and inelastic scattering) and propagation. Therefore, the present article intends to investigate the real component $\varepsilon_1(\omega)$ and imaginary component $\varepsilon_2(\omega)$ of dielectric function $\varepsilon(\omega)$, where $\varepsilon_2(\omega)$ exhibits absorption features of solids, directly related to the band structure of solids, whereas the real part $\varepsilon_1(\omega)$ is important in the study of reflectivity and loss function, etc. The reflectivity and loss function can be obtained from $\varepsilon_1(\omega)$ and $\varepsilon_2(\omega)$. The detail of optical properties calculations can be found elsewhere [93–95].

The metallic nature of the studied MAX compounds is confirmed from the DOS as shown in Fig. 3. Thus, it is scientifically significant to correct the intra-band contribution to the imaginary part of the dielectric constants. To increase the low energy part of $\varepsilon_2(\omega)$, a plasma frequency (3 eV) and a damping of 0.05 eV is introduced [96]. In addition, the obtained spectra have been broadened by the use of a Gaussian smearing (0.5 eV). The optical properties of $Hf_2AB$ (A = Pb, Bi) are calculated up to 20 eV for both [100] and [001] direction to showcase the anisotropic nature and presented in Fig. 7. The optical anisotropy is clearly demonstrated by figures presented here for both polarization directions.

Fig. 7 (a) shows the $\varepsilon_2(\omega)$ for both titled borides in which a clear difference is observed between the values for both directions. The both compounds exhibit almost identical spectra for [100] and [001] directions after 2.5 eV of incident photon energy. Fig. 7 (a) shows the $\varepsilon_1(\omega)$ for both titled borides that has a large negative values for both directions. This is the Drude-like nature that confirms the metallic nature of compounds, in good agreement with band structure results.

The most important optical constant of MAX phase materials from application point of view is the reflectivity that is calculated [Fig. 7 (c)] to assess the suitability of the titled borides as cover materials to prevent the solar heating. Li *et al.* [95,96] have predicted the MAX phase materials for the same based on the values of reflectivity, they predicted that the materials with reflectivity 44% can be used as coating materials to reduce solar heating. As seen from Fig. 7 (c) that both compounds fulfill the above conditions. Thus, they can be used for the same as discussed here. The values of *R*(0) for both [100] and [001] polarization directions of the electric field are 0.62 and 0.55 for $Hf_2PbB$; 0.66 and 0.51 for $Hf_2BiB$, respectively.

We have also calculated the loss function of $Hf_2AB$ (A = Pb, Bi) compounds as shown in Fig. 7 (d). As this function signifies the energy loss of electrons during the propagation trough the materials, therefore, the zero value of L ($\omega$) up to certain energy indicates no energy loss of electrons. The peak observed at certain energy depicts the character of plasma oscillation and defines the characteristic frequency that is known as the plasma frequency $\omega_p$ of the material [97]. At $\omega_p$, the real part of the dielectric function, $\varepsilon_1(\omega)$, becomes positive from below when the $\varepsilon_2(\omega) < 1$. The material exhibits dielectric response after $\omega_p$. The $\omega_p$ for both [100] and [001] polarization directions are 13.4 eV and 14.2 eV for the $Hf_2PbB$; 13.7 eV and 14.9 eV for the $Hf_2BiB$, respectively, where a rapid decline of reflectivity is noticed.

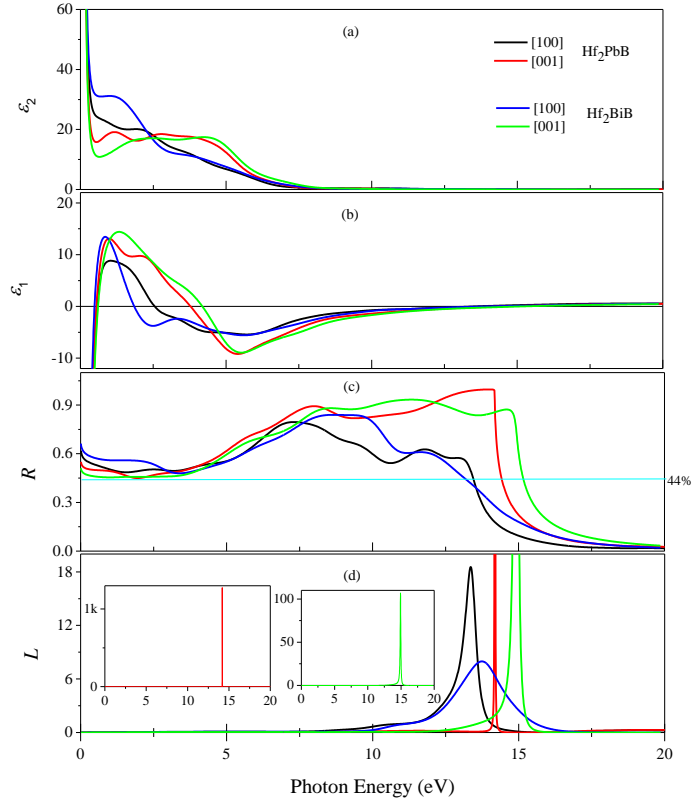

**Fig. 7** The (a) real ($\varepsilon_1$), and (b) imaginary ($\varepsilon_2$) part of dielectric function, (c) reflectivity (R) and (d) loss function (L) of Hf$_2$AB (A = Pb, Bi) MAX phase borides.

## 4. Conclusions

In the present study, we have performed DFT calculations to explore the mechanical behavior including their anisotropic nature, electronic, thermal and optical properties of Hf$_2$AB (A = Pb, Bi) MAX phase borides. The obtained lattice constants agree well with prior reported values. The titled borides are mechanically stable as confirmed from the stability criteria using stiffness constants. The mechanical properties of Hf$_2$AB (A = Pb, Bi) are higher than those of Hf$_2$AB (A = Ga, In) but smaller than those of Hf$_2$AB (A = Al, S). The Vickers hardness of the Hf$_2$BiB phase is higher than that of the Hf$_2$PbB phase that is in good agreement with lower energy position of the peaks in the DOS. The titled materials are brittle in nature like other MAX phases materials. The electronic band structure and DOS imply that these materials are metallic in nature and anisotropic electronic conductivity is confirmed from different energy dispersion along different direction. The studied materials shows highly anisotropic in nature by calculating different anisotropic indexes and presenting 2D as well as 3D of Young's modulus, linear compressibility, shear modulus and Poisson's ratio. The Debye temperature ($\Theta_D$) of Hf$_2$AB (A = Pb, Bi) are found to be lower than that of other Hf based 211 MAX borides. The obtained minimum thermal conductivity of Hf$_2$AB (A = Pb, Bi) is very low, comparable to other MAX phase materials. A low anharmonic effects within the studied borides is confirmed from the values of Grüneisen parameter. The studied reflectivity spectra confirmed their possible use as cover materials to diminish solar radiation.